\documentclass[a4paper,10pt]{article}
\usepackage{cite}
\usepackage[ref]{overcite}
\usepackage{graphicx}
\begin{document}

\twocolumn
\title{Plasmon-assisted quantum entanglement}
\author{E. Altewischer, M.P. van Exter, and J.P. Woerdman \\ 
Leiden University, Huygens Laboratory, P.O. Box 9504,\\ 2300 RA Leiden, The Netherlands}
\date{(March 8, 2002)}
\maketitle

\sloppy
{\bf The state of a two-particle system is called entangled when its quantum mechanical wave function cannot be factorized in two single-particle wave functions. Entanglement leads to the strongest counter-intuitive feature of quantum mechanics, namely nonlocality\cite{Greenberger,Bouwmeester}. Experimental realization of quantum entanglement is relatively easy for the case of photons; a pump photon can spontaneously split into a pair of entangled photons inside a nonlinear crystal. In this paper we combine quantum entanglement with nanostructured metal optics\cite{Raether} in the form of optically thick metal films perforated with a periodic array of subwavelength holes. These arrays act as photonic crystals\cite{Ebbesen,Martin,Ghaemi,Grupp} that may convert entangled photons into surface-plasmon waves, i.e., compressive charge density waves. We address the question whether the entanglement survives such a conversion. We find that, in principle, optical excitation of the surface plasmon modes of a metal is a coherent process at the single-particle level. However, the quality of the plasmon-assisted entanglement is limited by spatial dispersion of the hole arrays. This spatial dispersion is due to the nonlocal dielectric response of a metal, which is particularly large in the plasmonic regime; it introduces ``which way" labels, that may kill entanglement}.

The samples that we have studied are two 1~mm~$\times$~1~mm ``metal hole arrays", each comprising a 200~nm thick gold layer perforated with a square grid of 200 nm diameter holes spaced with a 700~nm lattice constant; a typical SEM picture is shown in the insert of Fig.~1. They were produced at DIMES (see acknowledgements) by first defining, with electron beam lithography, arrays of dielectric pillars on a 0.5~mm thick glass substrate, subsequently evaporating the gold layer onto a 2~nm thin titanium bonding layer, and finally etching away the pillars to leave the array of air holes. 

Transmission spectra of two hole arrays, measured with a spectrometer using normally incident white light, are shown in Fig.~1. One can clearly see the resonances due to excitation of Surface Plasmons (SP) on either of the metal-dielectric boundaries. At these resonances the measured transmission can be orders of magnitude larger than the value obtained from classical diffraction theory for subwavelength holes \cite{Ebbesen,Bethe}. In a simple picture, the surprisingly large transmission is due to the coupling of photon to SP on one side of the metal, subsequent tunnelling of the SP through the holes to establish a SP at the other side, and final re-radiation into a photon\cite{Martin}. Other prominent features in the spectra are the transmission minima associated with Wood anomalies \cite{Ghaemi}. At this moment the theoretical description of the full transmission spectrum is incomplete, but the role of the SP herein is well established\cite{Martin,Ghaemi,Grupp}. The resonance used in our experiment extends from 800-830 nm and is associated with the ($\pm$1,$\pm$1) SP mode on the glass-metal interface. Peak transmissions of the two arrays are typically 3\% (dashed curve) and 5\% (solid curve) at a wavelength of 813~nm; these values are much larger than the value of 0.55\% obtained from classical theory \cite{Bethe}. The difference in transmission between the two nominally identical hole arrays is ascribed to production imperfections. 

\begin{figure}[!ht]
    \centerline{\includegraphics[width=7.5cm, clip, trim= 0 25 0 0]{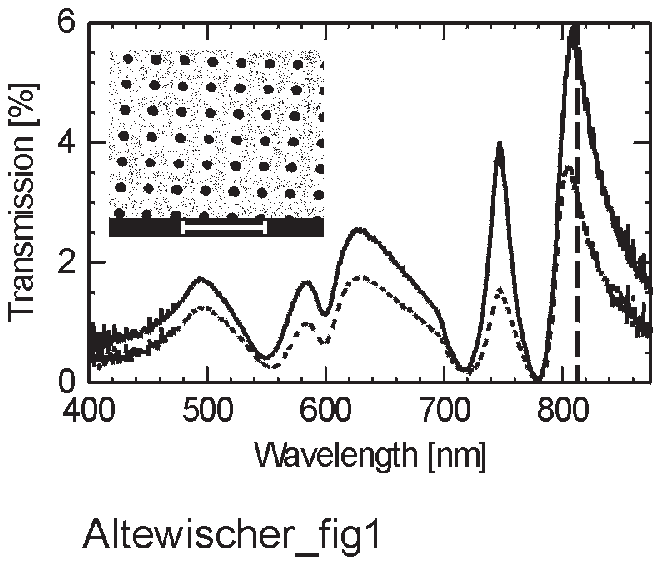}}
    \caption{\sffamily Wavelength-dependent transmission of the two hole arrays used in the experiment. The dashed vertical line indicates the wavelength of 813~nm used in the entanglement experiment. Inset: Scanning Electron Microscope picture of part of a typical hole array. The scale bar is 2~$\mu$m long.}
\end{figure}

As a mainly longitudinal, compressive electron density wave, a SP propagates along the polarization direction of the incident light, following a certain dispersion relation\cite{Hecht,Sonnichsen}. In order to confirm this for our samples we have measured polarization-resolved transmission spectra of one of the hole arrays for various angles of incidence $\theta _{inc}$ of plane-wave radiation (see Fig.~2). Angle tuning is expected to shift the various resonances in the transmission spectrum in different ways. As our interest is in a SP that propagates along the ($\pm$1,$\pm$1) direction, we have varied the angle of incidence by tilting the samples along the diagonal ($\phi = 45^\circ$) axis of the square hole array. For a polarization orthogonal to this tilting axis (Fig.~2a) the main peak at 810~nm splits and shifts for increasing $\theta _{inc}$; for a polarization along the tilting axis (Fig.~2b) the main peak remains at the same spectral position. These results confirm the relation between polarization and propagation and thus the 2D photonic band structure of the SPs.

\begin{figure}[!ht]
    \centerline{\includegraphics[width=8cm, clip, trim= 2 25 0 0]{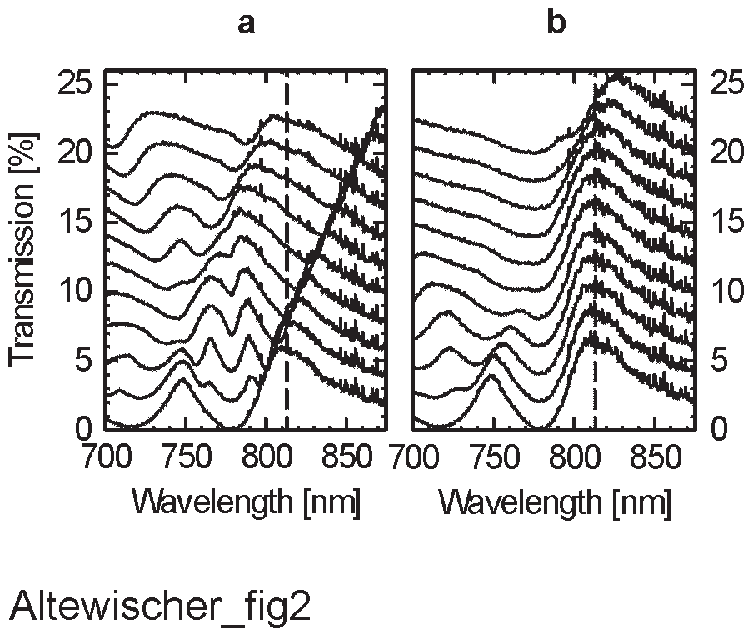}}
    \caption{\sffamily Wavelength-dependent transmission of a hole array for various angles of incidence at the two polarizations. The lowest curve is measured with the probe beam at normal incidence, while consecutive curves are measured at increasing angles of incidence (one degree steps) by subsequent tilting of the square hole array around a diagonal. These curves have been plotted with a subsequent vertical offset of 2\%. The dashed lines indicate the 813~nm wavelength used in the entanglement experiments. The resonance at this wavelength shows a complicated splitting for a polarization orthogonal to the tilting axis ({\bf a}), while it is stationary for polarization along this axis ({\bf b}).}
\end{figure}

We generate polarization-entangled photons with the standard method of type II Spontaneous Parametric Down-Conversion (SPDC) \cite{Kwiat,Kurtsiefer} depicted in Fig.~3. A 240~mW CW Krypton-ion laser beam at a wavelength of 406.7~nm is directed onto a 1~mm thick BBO nonlinear crystal, where the beam diameter is $\approx$~0.50~mm (full width at 1/e$^2$ points). Inside the nonlinear crystal a small fraction of the pump photons is down-converted into twin photons at the double wavelength; these are emitted along two intersecting cones. Polarization-entangled photon pairs are selected with pinholes at the crossings of these cones; the size of the pinholes (far-field diameter 5~mrad) was chosen as a compromise between high yield and good quantum entanglement. Lenses L of 40~cm focal length produce a one-to-one intermediate image of the pumped area, which will be used (see below) to accommodate the hole arrays A1 and A2. After passing through polarizers P1 and P2 the entangled twin photons are focused through interference filters (10~nm bandwidth centered at 813~nm) onto single-photon counting modules (Perkin Elmer SPCM-AQR-14). The photodetectors act as bucket detectors, i.e., they impose no further transverse mode selection (this is only done by the pinholes). The signals from the two detectors are combined electronically in a coincidence circuit with a time window of 2~ns, to provide the two-photon coincidence rate. 

\begin{figure}[!ht]
    \centerline{\includegraphics[width=8cm, clip, trim= 0 14 0 0]{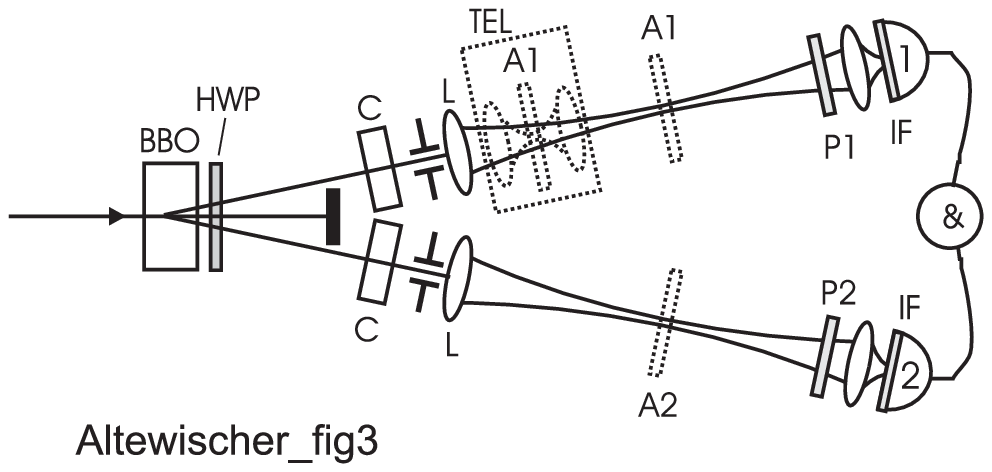}}
    \caption{\sffamily Experimental setup, comprising a BBO generating crystal, pinholes, 40~cm-focal length lenses L, polarizers P1 and P2, interference filters IF, single photon counters 1~and~2, and a coincidence circuit (2~ns). Beam walk-off is compensated for with the standard quantum eraser comprising a half-wave plate HWP at 45$^\circ$ and compensating crystals C with a thickness equal to half of that of the generating crystal\cite{Kwiat,Kurtsiefer}. The dotted objects are present only in some experiments; they show the hole arrays A1 and A2 at the image positions created by lenses L, or, alternatively, in the focus of the confocal telescope TEL (15~mm-focal length lenses).}
\end{figure}

In a simplified picture, the generated polarization-entangled state is
\begin{equation}
| \Psi \rangle = \frac{1}{\sqrt{2}} \, 
\left( \, |H_1 V_2 \rangle \,\,+\,\, e^{i\theta} \,|V_1 H_2 \rangle \, \right)\,\,,
\end{equation}
where the state $|H_1 V_2 \rangle$ represents the simultaneous emission and propagation of a $H$-polarized photon in beam~1 and a $V$-polarized photon in beam~2. The $H$- and $V$-directions are defined by the orthogonal birefringent axes of the generating BBO crystal and all spatial information is implicitly contained in the beam labelling. By tilting one of the compensating crystals (C in Fig. 3) the quantum phase $\theta$ can be set.

In a more accurate description, where the spatial information within the beams is properly accounted for, the two-photon state in a given transverse plane is
\begin{eqnarray}
| \Psi \rangle \propto \int {\rm d}\vec{q}_1 \, \int {\rm d}\vec{q}_2 \,\,\,\left\{ \Phi_{\rm HV} \left( \vec{q}_1,\vec{q}_2 \right) |H \vec{q}_1 \,; V \vec{q}_2 \, \rangle \right. \nonumber \\
\left.+\,\, e^{i\theta} \,\Phi_{\rm VH} \left( \vec{q}_1,\vec{q}_2 \right) |V \vec{q}_1 \,; H \vec{q}_2 \, \rangle \,\right\}, 
\end{eqnarray}
where the integration is over the transverse modes present in beam 1 and 2, labelled by wavevectors $\vec{q}_i$ ($i$=1,2). $\Phi_{\rm HV} \left( \vec{q}_1,\vec{q}_2 \right)$ is the two-photon probability amplitude for finding a H-polarized photon with wavevector $q_1$ in combination with a  V-polarized photon with wavevector $q_2$, and vice versa for $\Phi_{\rm VH} \left( \vec{q}_1,\vec{q}_2 \right)$. These functions are similar to the one normally used\cite{Atature,Saleh}, but now with polarization labelling and taken specifically at frequency degeneracy.  If the functions $\Phi_{\rm HV}$ and $\Phi_{\rm VH}$ are identical, i.e., in the absence of polarization-dependent components, they can be factorized out and Eq.~(2) reduces to the simple form of Eq.~(1). If the functions are different, polarization and spatial information are intertwined and the degree of polarization entanglement is reduced. Mathematically, it is the overlap integral between the two $\Phi$-functions in Eq.~(2) that determines the degree of entanglement, as it measures the spatial indistinguishability of the polarization-entangled photons\cite{Saleh}.

In the absence of the hole arrays our setup produces typically $3.2 \times 10^4$ coincidence counts per second, being $\approx$~25\% of the single count rate. A measure for the purity of the quantum entangled state is the so-called visibility of the biphoton fringe\cite{Kwiat,Kurtsiefer}. This visibility was typically $V_{0^\circ} = 99.3\%$ and $V_{45^\circ} = 97.0\%$ at polarization angles of 0$^\circ$ and 45$^\circ$, respectively (see Table). The high value at 45$^\circ$ shows that the natural preference of the generating BBO crystal for its birefringent axes (0$^\circ$ and 90$^\circ$) was almost completely removed in our setup by the compensating crystals, which act as quantum erasers.

Placement of the two hole arrays in the two ``beam lines'' of course leads to a dramatic reduction of single and coincidence counts. Coincidence count rates are typically 55~s$^{-1} $ at the optimum setting of the detecting polarizers, which is consistent with the transmissions of the arrays given above. We again performed a measurement of the purity of the entangled state and found that the visibilities were now $V_{0^\circ} = 97.1\%$ and $V_{45^\circ} = 97.2\%$, respectively. In Fig.~4 the corresponding fourth-order quantum interference fringes are shown for polarizer P2 fixed at 0$^\circ$ and 45$^\circ$, respectively, and P1 varying in steps of 10$^\circ$. These measurements show that the quantum entanglement survives the transition from photon to SP and back.

\begin{figure}[!ht]
    \centerline{\includegraphics[width=7cm, clip, trim= 0 14 0 0]{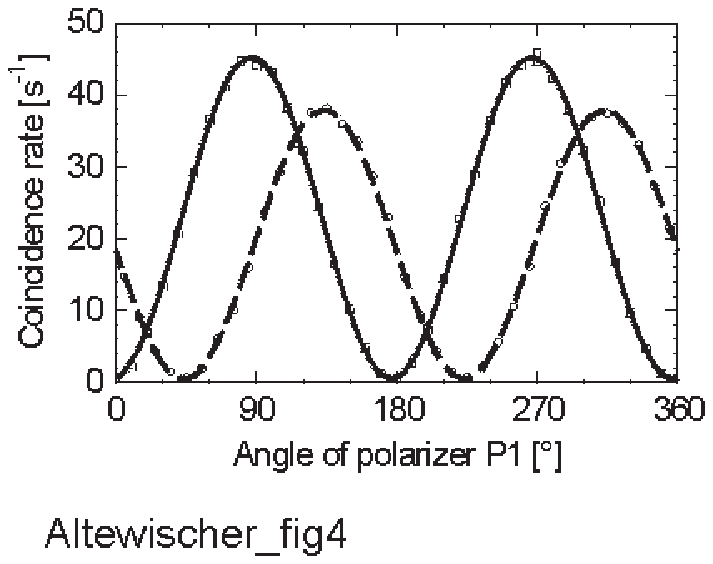}}
    \caption{\sffamily Biphoton fringes, corresponding to fourth order quantum interference, as measured with the two hole arrays in place, for P2 fixed at $0^\circ$ (solid curve) and $45^\circ$ (dashed curve) and P1 varying in steps of 10$^\circ$.}
\end{figure}

As a further demonstration of the SP-assisted quantum entanglement we performed a standard measurement of the so-called S-parameter, as described in ref. \cite{Kwiat}, on a singlet Bell state ($\theta=\pi$ in Eq.~1). This experiment, which took 16 runs of 100 seconds each, gave a value of $|S| = 2.71 \pm 0.02$, which is a violation of the inequality of Clauser, Horne, Shimony, and Holt \cite{CHSH}, by 35 standard deviations.  

Further experiments were performed on a setup with only a single array in one of the beams. The results for this case look very similar to those for the experiment with two arrays; again the entanglement was practically unaffected (see Table). This is to be expected as the two-photon wavefunction of Eq.~(1) is perturbed by changes in either of its single-photon components; in principle a single array can kill all of the entanglement. The difference between the two single-array experiments, as reported in the table, is due to imperfections in array 2, which are also observable in its (single-photon) polarization-dependent transmission. As the measurements using only array 2 gave results very similar to the situation with both arrays in place, these imperfections must be the key culprit for the somewhat limited visibilities in the two-array experiment. 

The most intriguing results of our single-array experiment are obtained when we focus the SPDC beam onto its hole array, using a confocal telescope (close to lens L) of two f~=~15~mm lenses symmetrically positioned around the array, as shown inside the dashed square in Fig.~3. Under these conditions we observe a drastic reduction of the degree of quantum entanglement: when the intra-telescope focus has a numerical aperture of 0.13~rad we observe visibilities of $V_{0^\circ} = 73 \%$ and $V_{45^\circ} = 87 \%$ (see table). 

\begin{table}
\begin{tabular}{l | @{\hspace{4mm}}l@{\hspace{4mm}} | @ {\hspace{4mm}}c@{\hspace{4mm}} | @{\hspace{4mm}}c@{\hspace{4mm}} }
Experiment & \hspace{4mm} R [s$^{-1}$] \hspace{4mm} & V$_{0^\circ}$ & V$_{45^\circ}$  \\
\hline 
No arrays & 32 k & 99.3 & 97.0 \\
Both arrays & 55 & 97.1 & 97.2 \\
Only array 1 & 1.6 k & 99.4 & 97.1 \\
Only array 2 & 1.0 k & 97.5 & 96.8 \\
Array 1, focussed & 1.1 k & 73 & 87 
\end{tabular}
\caption{\sffamily Biphoton fringe visibilities under various conditions.
Here R stands for the measured coincidence count rate, V$_{0^\circ}$ and V$_{45^\circ}$ for the measured visibility 
for one of the polarizers fixed at 0$^\circ$, 45$^\circ$ respectively.} 
\end{table}

The observed reduction in visibility upon focusing can be naturally explained as a consequence of spatial dispersion in the hole arrays. Spatial dispersion is a little-known phenomenon that refers to the fact that the optical constants of a medium depend on the {\it wavevector} of the incident light as a consequence of the non-local relation between the optical excitation in a medium and the incident optical field. It can lead, surprisingly, to polarization anisotropy of a cubic crystal, since the cubic symmetry is broken due to the finite size of the optical wavevector\cite{LL} (As an aside, we note that this effect has recently been demonstrated to have serious consequences for the use of high-precision crystalline optics for lithographic wafer production in the far ultraviolet\cite{Burnett}). Spatial dispersion is usually negligible as compared to frequency dispersion, i.e., the fact that optical constants depend on the {\it frequency} of the incident light. However, spatial dispersion can become significant in conducting media, where the motion of free charge carriers causes non-locality over distances much greater than atomic dimensions \cite{LL}. SPs are an extreme case thereof as these entities are not at all local but instead propagate along the dielectric-metal interface at nearly the speed of light over distances of many optical wavelengths \cite{Raether,Hecht,Sonnichsen}. 

As a result of this spatial dispersion, the near-field distribution of the photons that are reradiated at the backside of the hole array differs from the spatial profile of the ``polarization-isotropic" incident photons. From symmetry arguments we expect a four-lobed profile that is oriented either along the array axes or along the diagonals, where the latter applies to the ($\pm$1,$\pm$1) SP that we probe, as sketched in Fig.~5. The unpolarized overlap region of these profiles corresponds to the focused incident light, whereas the polarized elongations are a result of SP propagation. The presence of these polarized lobes introduces the possibility to distinguish the polarization of the photons on the basis of their spatial near-field coordinates. This will then automatically kill part of the polarization entanglement, just as any ``which way" information will do. A similar argument applies to the far-field, where the SP propagation creates a difference between the two $\Phi$-functions in Eq.~(2) after passage through the hole array. Note that the observed reduction in visibility is much stronger for $V_{0^\circ}$ than for $V_{45^\circ}$, contrary to what one generally finds without using hole arrays\cite{Kwiat,Kurtsiefer}. This observation is consistent with the fact that we excite SPs propagating in the ``diagonal" ($45^\circ$) directions.

\begin{figure}[!ht]
    \centerline{\includegraphics[width=4cm, clip, trim= 0 15 0 0]{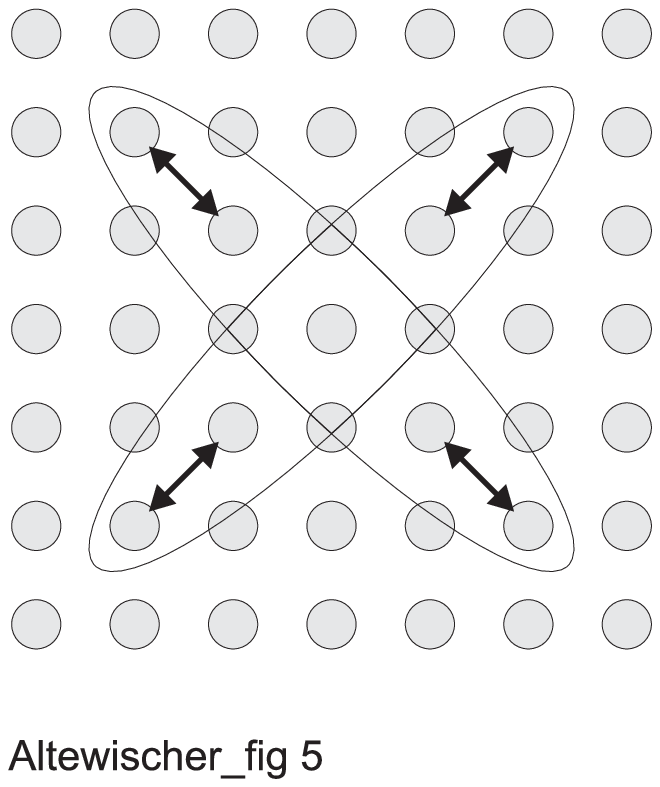}}
    \caption{\sffamily Schematic picture of the near-field at the back side of array~A1 when this is positioned inside the telescope.}
\end{figure}

The reduction of the visibility due to this effect is expected to be significant when the propagation length $l$ of the SP is comparable to the spot size of the focused incident beam, or rather its transverse coherence length. This is indeed the case for our experiment: based on the spectral width of the transmission peak at 810~nm, we estimate $l \approx 4~\mu$m; the number of holes ''covered" by the propagating SPs is thus about~10. In comparison, the spot size is $\approx$~8~$\mu$m ( =~0.5~{\rm mm}~$\times$~15/800) and the transverse coherence length (set by the 5~mrad far-field selection) is $\approx$~4~$\mu$m. We note that the propagation length differs by an order of magnitude from the value $l \approx 40~\mu$m predicted for a smooth pure gold film\cite{Raether}; we ascribe this difference to extra damping by the titanium bonding layer and the hole patterning.  

From a general perspective, the observed conservation of quantum entanglement for the conversion from photon $\rightarrow$ surface plasmon $\rightarrow$ photon is a first demonstration of the true quantum nature of SPs. All experiments on SPs so far do not probe their quantum nature, but only their semiclassical dispersion relation c.q. wave nature. In this perspective, it is interesting to note that the true quantum nature of the {\em photon} was only established in 1977, in anti-bunching experiments \cite{Lu,Kimble}. Furthermore, a simple estimate shows that SPs are very macroscopic, in the sense that they involve some $10^{10}$ electrons. Our experiment proves that this macroscopic nature does not impede the quantum behaviour of SPs, as they can be entangled with photons to yield the expected fourth-order interference. 

Based on the existing theory of spatial dispersion \cite{LL} the observed reduction in visibility in the focused setup is strongly linked to the square symmetry of the lattice and the ($\pm 1,\pm 1$) character of the SP that we have used. In the near future we plan to study hole arrays based on other Bravais lattices and space groups, of which there are 5, respectively 17 different types in 2 dimensions \cite{Kittel}. A very interesting case will be the hexagonal lattice, where the spatial dispersion is expected to be independent of the orientation of the lattice.

In conclusion, by addressing the topic of plasmon-assisted quantum entanglement we have combined two intriguing fields of research, being (i) quantum information, and (ii) nanostructured metal optics. We hope that our work will stimulate studies of entanglement transfer to other condensed-matter degrees of freedom. 

\footnotesize{

\vspace{0.5cm}
{\bf Acknowledgements} This work forms part of the program of FOM. We thank A. van Zuuk and E. van der Drift at the Delft Institute of Micro-Electronics and Sub-micron Technology (DIMES) in Delft, the Netherlands, for the production of the hole arrays. We thank G. Nienhuis for theoretical discussions.
\newline

\noindent Correspondence and requests for material should be addressed to E.A. (e-mail: erwin@molphys.leidenuniv.nl).}
\newpage

\end{document}